# Strain Induced Relaxor-type Ferroelectricity Near Room Temperature in Delafossite $CuCrO_2$

Preeti Pokhriyal[1,2], Ashok Bhakar[1,2], Parasmani Rajput[3], M. N. Singh[1], Pankaj Sagdeo[4], N. P. Lalla[5] , A. K. Sinha[1,2], Archna Sagdeo[1,2,a]

1. *Synchrotrons Utilization Section, Raja Ramanna Centre for Advanced Technology, Indore 452013, India*
2. *Homi Bhabha National Institute, Training School Complex, Anushakti Nagar, Mumbai 400094, India*
3. *Atomic & Molecular Physics Division, Bhabha Atomic Research Center, Trombay, Mumbai 400085, India*
4. *Material Research Laboratory, Discipline of Physics and MEMS, Indian Institute of Technology Indore, Simrol 453552, India*
5. *UGC-DAE Consortium for Scientific Research, Indore 452017, India*

a) Corresponding author: archnaj@rrcat.gov.in

Polycrystalline samples of $CuCrO_2$ were synthesized by solid state reaction method. Temperature dependent dielectric measurements, synchrotron x-ray diffraction (SXRD), pyroelectric current and Raman measurements have been performed on these samples. Evidences of the presence of relaxor type ferroelectricity, which otherwise have gone unnoticed in $CuCrO_2$ system (a member of delafossite family) near room temperature, have been presented. Presence of broad maximum in dielectric permittivity and its frequency dispersion indicates relaxor-type ferroelectricity in $CuCrO_2$ near room temperature. Careful analysis of temperature dependent SXRD data and Raman spectroscopic data indicates that the distorted $CrO_6$ octahdera, is giving rise to strain in the sample. Due to this strain, polar regions are forming in an otherwise non-polar matrix, which is giving rise to relaxor type ferroelectricity in the sample. Regularization of $CrO_6$ octahedra and disappearance of disorder induced peak in Raman spectra at high temperatures could be the reason behind observed dielectric anomaly in this sample. Present investigations propose that relaxor type ferroelectricity near room temperature is an inherent property of the $CuCrO_2$ system, making it a fascinating material to be explored further.

## I. INTRODUCTION

Transition metal oxides exhibit an abundance of fascinating physical properties covering an exceptionally wide range of phenomena in solid state physics.[1] Out of the list of these properties, multiferroicity is one of the properties that is currently in focus of material science. Multiferroics (MFs) are the class of materials that exhibit atypical cross-coupling between two ferroic orders, viz. ferromagnetic and ferroelectric, giving rise to potential technological applications as well as the underlying interesting physics.[2–4] Investigations on MFs has been intensified after the advent of spin driven (or type-II) multiferroics, where the magnetic ordering gives rise to the breaking of inversion symmetry.[5,6] Delafossites in this context, in the recent past have emerged as promising materials since they not only exhibit structural and magnetic transitions along with spontaneous spin-lattice coupling properties but also comprise of many interesting applications such as catalyst, luminescent materials, solar cells etc.[7–9]As these materials have combination of optical transparency and electrical conductivity, they have potential to be used in LCD displays, light emitting diodes etc.[10,11] They even show interesting phenomena such as negative thermal expansion[12] and transparent conductivity.[13–15] These materials are prototype of $ABO_2$ structure (forming layered structure), where A and B are monovalent and trivalent cations respectively. These structures consist of edge shared $BO_6$ octahedral layers. The A cation is linearly bonded to two oxygen atoms of upper and lower $BO_6$ layers, thus forming a linear O-A-O dumbell along c axis. Depending upon stacking of these layers delafossites can stabilize either in rhombohedral (R-3m) or hexagonal ($P6_{3/mmc}$) structure, both of which are centro-symmetric structures. Primitive rhombohedral cell consists of four atoms: A, B and two O atoms. However hexagonal cell is conventionally used to describe rhombohedral structures. In present case, A and B cations occupy 3a (0,0,0) and 3b (0,0,0.5) Wyckoff positions and O anions occupy 6c (0,0,u) positions.[16] Due to edge sharing geometry of $BO_6$ octahedra in the ab-plane, there is strong repulsion between trivalent B cations along shared edge, which thereby reduces O-O distance to contact distance.[17] These octahedra are flattened giving rise to two different O-O bond distances and hence deformed octahedra.[18] In addition to distorted octahedra, strong disorder in local structure has also been speculated in delafossites through Raman measurements.[19,20]

$CuCrO_2$ (CCO) is a member of delafossite family and have rhombohedral crystal structure with R-3m space group. Recently Seki *et al.* have shown that CCO exhibit multiferroicity at magnetic ordering temperature ~24K.[21] Around this temperature antiferromagnetic interaction take place between $Cr^{3+}$ ions that forms a triangular lattice in ab-plane. Ferro-electricity has been observed in magnetically ordered state without application of magnetic field due to proper screw magnetic ordering below Neel's temperature.[21] There are different mechanisms proposed for emergence of spin driven ferroelectricity.[22–24] Along with multiferroicity, it is also reported that CCO shows superconductivity at low temperatures (Tc~118K).[25] There has been a lot of interest among researchers to understand p-type conduction mechanism in this material. Recent theoretical work by Scanlon and Watson have suggested that $Cu^{1+}$-$Cu^{2+}$ hole mechanism is dominant than $Cr^{3+}$-$Cr^{4+}$ hole mechanism giving rise to p-type conduction. They also showed that the DOS at the valence band maxima is mainly due to Cu 3*d* states.[26] Experimentally, x-ray emission spectroscopy (XES) have also shown that the DOS at the top of the valence band in the Cu-based delafossites is governed by the Cu 3*d* character.[27]

Therefore, delafossites, in particular CCO, is an interesting material to be explored in many aspects. However, it must be emphasized here that, experimental and theoretical studies, principally dielectric properties reported on this system are very few. Also, studies are mostly confined to investigate magnetic properties of this system and have mainly been restrained to low temperatures, in the immediate vicinity of the magnetic transition temperature ~24K, where anomaly in the dielectric constant has also been observed.[21,28–30] Apart from this, it has been reported that another member of delafossite family, $CuFeO_2$, exhibit colossal dielectric permittivity well above the magnetic transition. Thus, it was revealed that this delafossite is very fascinating to be explored, as it shows multiferroicity at low temperatures (~14K) and colossal dielectric permittivity at high temperatures (~200K-400K).[31]

Keeping all this in mind, it is anticipated that CCO being member of delafossite family, which also show multiferroicity at low temperature, is expected show interesting dielectric properties in vicinity of room temperature and at high temperatures, as is the case with $CuFeO_2$. Therefore, in the present study, it is deliberated to explore temperature dependent dielectric properties of CCO in the vicinity and above room temperatures which is well above the magnetic transition temperature. For this, in the present work, dielectric properties of CCO system have been explored in the immediate vicinity of room temperature. To support these findings of dielectric measurements, SXRD measurements and Raman studies have been carried out in the similar temperature range.

## II. EXPERIMENTAL DETAILS

Polycrystalline sample of CCO was synthesized via solid state reaction method. Stoichiometric amounts of high purity raw materials $Cu_2O$ (99.9%) and $Cr_2O_3$ (99.9%) were taken and grounded thoroughly in liquid medium (Propanol). Dried mixture was heated at 1473 K for 12 hours in air with a heating rate of 5K/min. Calcined powder was mixed with binder, and pressed into pellets followed by sintering at 1200°C for 24 hours. Phase purity of the samples were confirmed by performing X-ray diffraction (XRD) measurements on Bruker D8 Advance diffractometer using laboratory x-ray source with Cu Kα radiation (wavelength

~1.54 Å), having a LYNXEYE detector. Measurement was carried out in Bragg- Brentano geometry in 2θ range of 10 to 110° with step size of 0.01 and with the scan speed of 1.0 sec/step. XRD pattern was analyzed by performing Rietveld refinement using FullProf program.[32] High resolution and high temperature SXRD measurements have also been performed using ADXRD beamline[33] BL-12, at Indus-2 synchrotron radiation source, at Raja Ramanna Centre for Advanced Technology, Indore, India. Measurements were performed on six circle diffractometer (Huber 5020) using Mythen detector. Mythen detector is one dimensional position sensitive detector consisting of 1280 channels. Resistive heater was used for heating the sample. Measurements were performed in θ-2θ mode. The X-ray wave length used in the present study is 0.756 Å and was accurately calibrated by measuring XRD pattern of NIST $LaB_6$ standard. X-ray absorption near edge spectroscopy (XANES) measurements at Cr (5989eV) and Cu (8979eV) K edge have been performed at Scanning EXAFS beamline BL-09[34], in transmission mode at Indus-2, synchrotron radiation source, at Raja Ramanna Centre for Advanced Technology, Indore, India. The incident and transmitted intensities were measured using gas filled ionization chamber detectors. XANES spectra for Cu and Fe foils were also measured in the same geometry for energy calibration. WAYNE KERR 6500B impedance analyzer has been used for temperature dependent (100K-480K) dielectric measurements in the frequency range of 3kHz to 110kHz. Temperature dependent measurements were performed using CRYOCON 22C temperature controller. Temperature of the sample was varied from 100K to 480K at a rate of 2K per minute. Silver paste was used for making electrical contacts on samples. Complex impedance analysis was performed by least-mean-square analysis using EC lab software. Raman measurements have been carried out using LABRAM HR dispersive spectrometer equipped with a 633nm excitation laser source and a CCD detector in backscattered mode. For temperature dependent measurements THMS600 stage from Linkam having accuracy of 0.1K is used. Temperature dependent pyroelectric current measurement was performed on sample using 6517B, Keithley electrometer.

## III. RESULTS AND DISCUSSIONS

### A. X-ray Diffraction

Figure 1(a) shows x-ray diffraction (XRD) pattern of the prepared CCO sample. No additional peaks are observed in the XRD pattern, confirming the phase purity of the prepared sample. The schematic of crystal structure is shown in figure 1(b) drawn using

VESTA software. The Rietveld analysis of XRD pattern reveals that as-synthesized sample crystallizes in rhombohedral crystal structure (space group R-3m). Structural parameters obtained from refinement are, a = b = 2.9748(2) Å, c = 17.1027(16) Å and volume of unit cell = 131.078(2) $Å^3$, which is consistent with the reported result.[25]

Since, in delafossites, lattice constant along c-axis is usually large as compared to the lattice parameter along other axis, it is expected that the peak intensity corresponding to the reflections along c-axis (for e.g. peaks corresponding to (00l) reflections) will also be larger than calculated values. Although, this unusual intensity variation has been observed in the XRD patterns of different reports from literature, to best of our knowledge, this intensity variation is attributed to preferred orientation without complete analysis. In the present investigation, during refinement of the structure, preferred orientation along this c-axis has also been considered. Although the fitting has improved for (006) peak with the consideration preferred orientation, still there is a considerable mismatch between the observed and calculated intensity of this peak. Apart from this, anisotropic broadening was also observed in few diffraction peaks (for e.g. peaks corresponding to (101), (012), (104), (018), (110), (1010), (116), (202) reflections etc.). Microstrain maybe one of the possible reasons for such observed anisotropic broadening. In order to account for that, along with preferred orientation, anisotropic broadening was also considered for refinement of the observed pattern. Stephen's broadening model[35] was used to take into account anisotropic broadening. For rhombohedral system in hexagonal setting, Stephen's model allows three anisotropic broadening parameters $S_{400}$, $S_{004}$ and $S_{112}$. Magnitude of parameters determines hkl dependent broadening. With this consideration, it has been observed that the quality of fitting is improved significantly and refined XRD pattern is shown in figure 1(a). The model considering preferred orientation and anisotropic microstrain, fits the data and the (006) & (104) peaks very well. Insets α and β of figure 1(a) and shows a comparison of the fitting considering without and with anisotropic strain for peak (006). Similarly Insets γ and δ of figure 1(a) and shows a comparison of the fitting considering without and with anisotropic strain for peak (104). Improvement in the fitting confirms the validity of the chosen model. Details of the fitting considering anisotropic broadening and the fitting parameters extracted will be discussed in the temperature dependent x-ray diffraction measurement section.

## B. XANES Measurements

XANES measurements at Cr K-edge and Cu K-edge have been performed in order to find out the oxidation states of Cr and Cu respectively. XANES spectra of sample along with their reference compounds $Cr_2O_3$, $CrO_3$ for Cr K-edge and $Cu_2O$, CuO for Cu K-edge have been shown in figure 2 (a) and (b), respectively.

### 1. Cr K- edge

XANES spectra at Cr K-edge shown in figure 2(a), indicates that edge position of CCO matches well with $Cr_2O_3$ which confirms that Cr cation is present in +3 oxidation state. It is well known that in $Cr^{3+}$ compounds, 3d states get splitted into two levels by octahedral crystal field splitting, which is reflected in the pre-edge features of the XANES spectra of these compounds.[36,37] XANES spectrum at Cr K-edge clearly shows two pre-edge peaks marked as features A and B, in the in-set of figure 2(a). Peaks A and B correspond to $1s\rightarrow 3d$ transitions into $t_{2g}$ and $e_g$ states respectively.[36,38] These transitions are quadrupole transitions taking place due to mixing of Cr(3*d*) and O (2*p*) states. It is important to note here that 3d-2p hybridization is only prominent when octahedra are distorted. Hence, it can be perceived that for CCO sample $CrO_6$ octahedra are also distorted. Crystal field splitting between $t_{2g}$ and $e_g$ states is found to be 3.2eV. This separation (splitting) between the peaks, agrees well with reported octahedral crystal field splitting between $t_{2g}$ and $e_g$ levels for several $Cr^{3+}$compounds.[39,40] Shoulder marked as C corresponds to 1*s* to 4*s* transition. This transition is symmetry forbidden, but appears when orbital mixing is present.[41,42] In CCO, Cr is present in high symmetry octahedral environment due to which shoulder C is very weak. Region D corresponds to main edge ($1s\rightarrow 4p$) transition. All these features are in agreement with the reported results for $Cr^{3+}$ compounds.[37]

### 2. Cu K-edge

XANES spectra at Cu K-edge shown in figure 2(b) indicates that Cu K-edge position in CCO, is slightly shifted towards CuO standard sample compare to standard $Cu_2O$ sample. Thus, a lack of perfect overlapping of Cu K-edge position in CCO with $Cu_2O$ indicates that Cu in CCO is not in exact $Cu^{1+}$ state, however a shift towards CuO suggests that small amount of $Cu^{2+}$ is also present in CCO. As $Cu^{2+}$ fraction is very small, therefore the pre-edge peak is not observed here. It is well known that $Cu^{1+}$ compounds show splitting of 4*p* level in the presence of ligand field.[43] For linearly coordinated $Cu^{1+}$ compounds due to repulsive interaction along ligand and metal bonds, $4p_z$ has higher energy than $4p_{xy}$ after splitting. In CCO system Cu is linearly coordinated with two oxygen ions. Ligand field of oxygen ions leads to splitting of 4*p* level. Thus, as shown in the inset of figure 2(b), peak A corresponds to $1s \rightarrow 4p_{xy}$ transition, B corresponds to $1s \rightarrow 4pz$ and C corresponds to transition from core level to continuum. These results are consistent with the results reported for the Copper based compounds.[43–45]Separation between peaks A and B was found ~11eV, which matches well with literature reported for $Cu^{1+}$ compounds.[38] Thus, from the XANES measurements of CCO it has been observed that Cr is present in $Cr^{3+}$ oxidation state as expected, however, Cu is present in mixed oxidation state. Majority of the Cu is in +1 oxidation state along with a small amount of +2 oxidation state.

### C. Dielectric measurements

Figure 3 (a) and (b) show the temperature dependent dielectric permittivity (ε') and the corresponding loss tangent (tanδ) data of CCO sample respectively. The measurements have been performed in the temperature range of 100-480K and in the frequency range of 3kHz-110kHz. However, for lucidity, the data have been shown for few selected frequencies. It can be observed that at low temperatures, ε' is small (~90). Low value of dielectric constant at low temperatures has been attributed to freezing of dipoles. When temperature reaches ~200K, a step-like increase in ε' is observed which reaches values greater than $10^3$. Corresponding to this rise in ε', peak in loss tangent is clearly visible. This peak in loss tangent data shifts to high temperature with increase in frequency (marked as region I), indicating presence of thermal relaxation in this temperature range. In order to understand this thermally activated relaxation mechanism in the sample, further analysis of the data has been performed within the realm of Arrhenius law ($f_p = f_0 \exp(-Ea / k_B T)$), where $\ln(f_p)$ is plotted as a function of 1000/T (inset (β)). Here $f_p$ is the peak frequency, $f_0$ the pre-exponential factor, $E_a$ the activation energy, $k_B$ the Boltzmann constant, and T is absolute temperature in kelvin. Activation energy for this relaxation process has been calculated, and is found to be 0.32eV. Such value of activation energy, corresponds to small polaronic hopping.[46] This result is consistent with outcome of XANES measurements that Cu is in mixed oxidation state of $Cu^{1+}$ and $Cu^{2+}$. Thus, hole hopping between $Cu^{1+}$ and $Cu^{2+}$ gives rise to p-type conduction.

With further increase in temperature, around 445K (at 3 kHz), a broad and smeared peak in ε' appears, as observed in the figure 3(a), and marked as region II. This peak shifts to higher temperature with increase in frequency. However, peak position (~387K at 3kHz) in the loss tangent data, corresponding to this peak in ε' data, remains almost at constant temperature with increase in frequency. The appearance of broad peak in ε' with a large value (for e.g. at 3kHz it is ~$10^4$), gives an indication of the presence of ferroelectric to paraelectric phase transition in this material. However, the peak observed in dielectric permittivity is not as sharp as expected in normal ferroelectrics, but is broad and smeared and also shows frequency dispersion, indicating that the sample is not normal ferroelectric.[47–49] Another possibility is of relaxor ferroelectric, where deviation from Curie-Weiss law is generally observed. Figure 4 (a) shows variation of inverse of permittivity (1/ε') as a function of temperature at frequencies 3kHz, 10kHz, 30kHz and 110kHz. Region fitted to Curie-Weiss law is shown by red line. The deviation from ideal Curie-Weiss law is clearly evident in figure 4(a), which further supports the speculation that the sample is not normal ferroelectric. Degree of deviation from the Curie-Weiss law can be calculated as:

$$\Delta T_m = T_{cw} - T_m \text{........(2)}$$

Where $T_{cw}$ is the temperature at which deviation in permittivity starts from ideal Curie-Weiss law and $T_m$ is the temperature at which maximum permittivity appears. At 10 kHz, temperature at which deviation from ideal Curie-Weiss law starts, is calculated to be 461.56K determined from Curie-Weiss fit and the temperature where maximum in permittivity appears is ~ 453.14K. Thus, $\Delta T_m$ is calculated to be 8.42K. To further explore the possibility of relaxor-type ferroelectricity, modified Curie–Weiss law is used, which has been proposed by Uchino and Nomura[50], to describe the diffuseness of the phase transition. This is defined as

$$\frac{1}{\varepsilon} - \frac{1}{\varepsilon_m} = \frac{(T-T_m)^{\gamma}}{C} \text{..........(3)} \quad , \quad 1 \leq \gamma \leq 2$$

Where C is constant and γ is the parameter which represents nature of phase transition. The value of this parameter is 1 for real ferroelectric and 2 for perfect relaxor ferroelectric.[51] Figure 4(b) shows variation of $\ln(1/\varepsilon - 1/\varepsilon_m)$ versus $\ln(T-T_m)$. From the slope of linear fit to the curve, value of ‘γ’ has been calculated to be 1.76 at 10 kHz, indicating that CCO has relaxor- type behavior. The value of ‘γ’ is consistent with reported literature for relaxor type ferroelectrics.[52]

Following the results obtained from above analysis it can be concluded that for CCO material, dielectric permittivity obeys Curie-Weiss law only at temperatures much higher than $T_m$. Large deviation from the Curie-Weiss law suggests that CCO is relaxor-type ferroelectric.

For ferroelectric materials it has been reported in literature that, they exhibit positive temperature coefficient of resistance (PTCR) above Curie temperature .[53,54] $BaTiO_3$, which is an insulator at room temperature, becomes semiconductor when doped, displays PTCR effect. Out of various models Heywang-Jonker model was found to be widely accepted in explaining this PTCR behavior.[53–55] According to them, when two grains of different orientation are in contact, their Fermi levels come in equilibrium, as a result space charge is developed at the boundaries between grains. Thus, a potential barrier is formed constituting these space charges.[56] Above Curie temperature ($T_c$), the surface states present at grain boundaries trap electrons. These trapped electrons at grain boundaries forms Schottky type potential barriers. Above $T_c$, with decrease in permittivity, potential barrier height increases and hence gives rise to an increase in resistivity. However, below $T_c$, because of high permittivity, barrier height is small, giving rise to reduction in potential barrier and hence drop in resistivity is observed. At transition, anomaly begins to start. This increase in resistance with increasing temperature stops when Fermi level has reached energies of donor impurities. After this point, materials regains its normal semiconducting behavior.[56] The extent of PTCR effect i.e. the change in the magnitude of resistance across $T_c$, depends on the processing conditions and hence the chemical composition of the grain and grain boundaries present in the sample. Thus, in principle PTCR effect must also be observed across the transition in relaxor type ferroelectrics. Therefore, in order to examine the presence of PTCR in this sample, variation of resistance as a function of temperature has also been explored. Figure 5 shows temperature dependence of measured resistance at different frequencies. A systematic decrease in resistance is observed with increase in temperature indicating negative temperature coefficient of resistance, which is a typical feature of semiconducting materials. However, above 400K, an increase in resistance with increasing temperature is observed upto 470K, after which again resistance begins to decrease with increasing temperature and CCO regains its normal semiconducting behavior above 470K. Thus, around this temperature range, CCO system shows insulating-like to metal-like transition.[57]

Thus, with this background, it is proposed that the presence of PTCR effect, in the present case, corroborates with the speculation of the presence of ferroelectricity in the CCO sample.

In, order to further confirm the presence of ferroelectricity, ferroelectric hysteresis loop measurements have also been attempted at room temperature for this sample. No hysteresis loop has been observed. However, the non-observance of ferroelectric hysteresis loop does not directly indicate absence of ferroelectricity in the sample. This has already been reported in oxygen vacancy doped $BaTiO_3$ samples.[58,59] It has been clearly shown that a ferroelectric hysteresis loop cannot be measured in metallic and semiconducting samples, since with the application of external field; flow of electric current gets induced instead of polarization switching. In such cases where ferroelectric hysteresis loops cannot be measured due to high leakage in the sample, presence of polarization through pyroelectric current measurements is generally shown.[60–63] Thus, temperature dependent pyroelectric current measurements have been performed on CCO sample.

## D. Pyroelectric current measurements

Pyroelectric current measurements were performed on poled CCO sample as a function of temperature using electrometer (6517 B, Keithley). An electric-field of 1.4kV/m was applied at 430 K (well above the temperature corresponding to maximum in dielectric permittivity) in order to align (pole) the ferroelectric regions. The sample was cooled down to 300K. After reaching 300K, the poling-field was switched off and the electrodes were shorted for about 30 minutes to completely eliminate the unwanted stray charges at electrodes (if any). Finally, pyroelectric current as a function of temperature was measured during heating cycle with heating rate of 3 K/min. An anomaly in the inset (α) of figure 6 is observed around 384K in the as measured pyroelectric current data, indicated by arrow. Figure 6 shows pyroelectric current as a function of temperature after background subtraction. Background subtraction was performed in order to separate out the required signal of pyro-current due to dipole polarization. The background of thermally stimulated current arises mainly due to release of leakage charge carriers, which get trapped at defect sites during poling process.[64] After background subtraction the peak due to polarization current can be clearly seen in figure 6 around 384K. Figure shows zero value of current in temperature range of 320K to 370K, which therefore confirms that peak around 384K, is not due to any artifact corresponding to leakage current. Thus, the existence of peak in pyroelectric current around the same temperature where peak in loss tangent data has been observed further supports presence of relaxor type ferroelectricity in this system near room temperature.

It should be noted at this point that CCO is reported to be a centro-symmetric (R-3m) structure, where in principle ferroelectricity should not be observed. Thus, to find out the origin of observed relaxor type ferroelectricity in this centrosymmetric system high resolution, SXRD measurements at room as well as high temperatures have been performed using Indus-2 synchrotron source. Temperature dependent Raman measurements have also been utilized in order to support the experimental inference.

### E. Temperature dependent XRD measurements

In order to study structural changes across the transition that are giving rise to observed relaxor type ferroelectricity, in an otherwise centrosymmetric CCO sample, high resolution synchrotron x-ray diffraction (SXRD) measurements were performed from room temperature up to 610K. Diffraction patterns at representative temperatures are shown in figure 7(a). For the sake of clarity, the patterns at different temperatures are vertically shifted. As can be observed from the figure, the diffraction patterns at different temperature appears to be similar, except for the peak shift, which is expected due to thermal expansion. This is shown by red dashed line in figure 7(b). However, it must be noticed here that at around $2\theta=46.9^{o}$, in the room temperature (RT) data, there are two peaks corresponding to Miller indices (0 0 18) and (2 1 4), as shown in figure 7(b) (where only these peaks are magnified). These two peaks merge as the temperature is increased and becomes single peak at ~473K. A considerable increase in the peak intensity at this temperature can also be observed. The merging of peaks and also the increase of peak intensity gives a clear indication of the change in structure of the system increasing its apparent order with the temperature, as will be discussed below in terms of bond length and bond angles. Nonetheless, the changes that are visible is restricted to just one peak, hence, it is speculated that the structure of the sample as a whole is not getting transformed, but some local changes in the structure are taking place. In order to extract the information about the local changes in the structure, thorough analysis of temperature dependent SXRD data by Rietveld refinement has been performed assuming the rhombohedral structure. The extracted lattice parameters from the refinement are shown in figure 8 (a) to (d).

A systematic increase in the lattice parameters ‘a’ and ‘c’ as well as in volume of the unit cell is observed, as is expected due to thermal expansion. However, a decrease in the c/a ratio is observed with the increase in temperature, indicates that the rate of increase of ‘a’ is larger than the rate of increase of ‘c’. Apart from this, anomalies in the lattice parameters ‘a’ and ‘c’ are clearly visible at around 400K and 450K. The derivative of the same shows peaks corresponding to the anomalies, as shown in the inset of figure 8(a) and (b). These anomalies are also evident in c/a ratio as well (shown in figure 8(c)), at the same temperature. In order to further understand this, various bond lengths, Cu-Cu, Cr-Cr, Cu-Cr, Cu-O and Cr-O have been extracted using Vesta software. These bond lengths are plotted in the figures 9 (a) to (d). It can be observed that the Cu-Cu and Cr-Cr bond lengths follows the trend of lattice parameter 'a', however, the Cu-Cr bond length, follows exactly the 'c' parameter. This is expected since the bond lengths Cu-Cu and Cr-Cr, lies in the ab-plane, whereas, bond length Cu-Cr lies along c-axis. Although, temperature dependence of these parameters gives an indication that the peak observed in the dielectric permittivity data is associated with the changes in the structure, nevertheless origin of correlation is not clearly evident from these parameters. Thus, the temperature dependence of various bond angles was also extracted. In delafossites bond angles $O_{in}$-Cr-O and $O_{out}$-Cr-O are not right angles (as expected for the ideal octahedra), instead they are different from $90^o$. Angle $O_{in}$-Cr-O is lower whereas $O_{out}$-Cr-O is greater than $90^o$. This is shown schematically in figure 10. This indicates that $CrO_6$ octahedra in CCO are distorted. This has also been reported in literature and hence the observation of the present study is consistent with the reported results.[12] This is also supported by the XANES measurements at Cr-K edge, as discussed above, where the appearance of pre-edge peaks indicate the presence of distortion in $CrO_6$ octahedra.

Variation of these two angles as a function of temperature is shown in the figure 11(a). It can be observed that bond angle $O_{in}$-Cr-O is decreasing whereas $O_{out}$-Cr-O is increasing with temperature which signifies that both these angles tends to approach towards $90^o$. This suggests that the distortion present in the octahedra is getting reduced with the increase in temperature. In delafossites, degree of distortion of octahedra can be estimated by calculating ratio O-$O_{in}$/O-$O_{out}$ as suggested by Doumerc *et al.*[18] O-$O_{in}$ is the distance two oxygen atoms of $CrO_6$ octahedron edge, parallel to ab-plane and O-$O_{out}$ is the length of octahedron edge delimited by two oxygen atoms of two consecutive oxygen planes on each side of $Cr^{3+}$ layer. Figure 11(b) shows variation of ratio O-$O_{in}$/O-$O_{out}$ i.e. degree of distortion, with increasing temperature. It can be perceived that distortion of octahedra decreases systematically with increasing temperature indicating octahedra is getting regularized at high temperatures. Along with this, it has been observed in the high temperature SXRD data, that (0018) and (214) peaks are getting merged with increase in temperature. Thus, it can be speculated that merging of peaks at high temperatures is due to reduction of distortion in $CrO_6$ octahedra.

Apart from this, it should also be noted that these oxygen atoms which are changing the $O_{in}$-Cr-O & $O_{out}$-Cr-O bond angles are connected to the Cu atoms, as shown in figure 10. Hence, the bond length Cu-O is getting shortened in the process (figure 9(c)). Thus, the changes in the $O_{in}$-Cr-O and $O_{out}$-Cr-O bond angles are thereby responsible for contraction of Cu-O bond lengths as shown in figure 9(c).

Also it should be noted here that, preferred orientation along with the anisotropic peak broadening model have been considered during the Rietveld refinement of SXRD data in order to fit the intensity of (006) peak in the observed pattern. It has been observed that out of the three anisotropic broadening parameters $S_{400}$, $S_{004}$ and $S_{112}$, only the parameter $S_{112}$ shows a systematic variation with temperature, and is shown in figure 12. It has been observed that the parameter $S_{112}$ decreases with temperature. The decrease in this parameter, suggests that the anisotropic strain in the sample, is decreasing with increase in temperature. Thus, it may be interpreted that the distorted octahedra is giving rise to the strain in the sample, both of which are decreasing with temperature. It has been reported in literature that presence of strain in the sample are many a times responsible for formation of nano regions, which might be polar in nature in otherwise non-polar matrix giving rise to relaxor type ferroelectricity.[47] It may be speculated that for present case also, polar nano regions formed due to presence of strain in the sample are responsible for observed relaxor type ferroelectricity. Raman spectroscopy is a well known technique to probe local disorder in any system, thus, to further support and corroborate the finding, temperature dependent Raman spectroscopy measurements have also been performed.

### F. Raman Spectroscopy measurements

CCO a member of delafossite family belongs to point group (-3m) $C_{3V}$ and space group R-3m. It has one formula unit per primitive cell, which give rise to 12 normal modes of vibration in the zone center[65],

$$\Gamma = A_{1g} + E_g + 3A_{2u} + 3E_u \quad .....(6)$$

Modes with symmetry 'A' are non-degenerate, while modes with 'E' symmetry are doubly degenerate. There are three acoustical and nine optical modes. Modes denoted by subscript 'g' are Raman active and modes denoted by subscript 'u' are infrared active, which also includes acoustic modes. Group theory calculations yield two Raman active modes for delafossites having $A_{1g}$ and $E_g$ symmetries.[66] 'A' modes corresponds to vibration of Cu-O bonds parallel to c-axis, while doubly degenerate 'E' modes correspond to vibration in triangular lattice perpendicular to c-axis.[67] It is well known that for centro-symmetric crystals normal modes will have definite parity. Odd modes depicted by 'u' subscript are infrared active $A_{2u}$+$E_u$. In acoustic modes both oxygen atoms should move in phase, while in Raman active modes they move out of phase.[65] Copper and Chromium are at rest in Raman active modes, only oxygen atoms vibrate. Evolution of Raman spectra in the temperature range from RT to 573K is shown in figure 13. For the sake of clarity the spectra are shifted vertically. It can be observed from the Raman spectra that along with Raman active modes, some additional modes are also present. These modes have been observed for poly crystalline[20,68] as well as single crystalline samples of CCO.[19] At room temperature modes are observed at 101cm$^{-1}$, 208cm$^{-1}$, 457.5cm$^{-1}$, 538cm$^{-1}$,623cm$^{-1}$ and 709.2cm$^{-1}$ wave numbers. Raman active modes $E_g$ and $A_{1g}$ appear at 457.5cm$^{-1}$ and 709.3cm$^{-1}$ respectively. $E_g$ and $A_{1g}$ symmetries are assigned to these modes on the basis of ab-initio calculations[66] and polarization dependent Raman measurements reported in literature.[66] Mode appearing around 208cm$^{-1}$ has been attributed to $A_g$ mode.[68] Peak appearing around 538cm$^{-1}$ correspond to vibrations of $CrO_6$ octahedral modes of delafossite structure.[69,70] These additional modes which are infrared active are not allowed according to Raman selection rules. In literature appearance of these modes in Raman spectra has been attributed to presence of Cu vacancies, interstitial oxygen atoms or tetrahedrally coordinated $Cr^{3+}$ on the Cu site. Theoretical calculation on other delafossite $CuAlO_2$ and $CuGaO_2$ also indicates presence of these modes.[65,66] There is no splitting of peaks corresponding to these Raman active modes $E_g$ and $A_{1g}$. Thus, it can be concluded that none of the Raman modes is driving the diffuse phase transition in CCO as indicated by the temperature dependent SXRD measurements. From XRD measurements it has been observed that $CrO_6$ octahedra are distorted. Two angles $O_{in}$-Cr-O and $O_{out}$-Cr-O depart from 90 degree, which is common feature of delafossites due to edge shared octahedral arrangement.[12] Presence of forbidden modes appearing in Raman spectra indicates lowering of local symmetry. This symmetry lowering might be due to octahedral distortion. With the increase in temperature these modes are disappearing, indicating decrease in disorder. In literature it has been shown that distortion in octahedra can develop local polarization in average centro-symmetric structures also.[71] Presence of forbidden bands in Raman spectra corresponding to Cr-O vibrations have been observed in the present case. Upon increasing temperature, mode appearing at 538cm$^{-1}$ reduces in intensity while mode at 623 cm$^{-1}$ initially gets weaken and then finally disappear around 393K, indicating regularization of octahedra with increasing temperature. This observation is consistent with the temperature dependent XRD results.

Thus, it has been found that CCO being globally centrosymmetric at room temperature indexed well (XRD) with space group R-3m, shows relaxor type ferroelectricity. The presence of which is getting masked due to increased large leakage current in the sample.

## IV. CONCLUSION

In the present investigation, single phase CCO sample have been successfully synthesized by solid state reaction method. Phase purity has been confirmed by x-ray diffraction measurements. XANES measurements confirm that Cr present in the sample is in +3 oxidation state while majority of Cu is present in +1 oxidation state. Appearance of dielectric anomaly and its frequency dispersion give indication of presence of relaxor type ferroelectricity in CCO system at room temperature. Careful analysis of temperature dependent XRD and Raman measurements reveals that local disorder is present in the sample. It has been speculated that distorted $CrO_6$ octahedra is giving rise to strained regions, which might be polar in nature, are responsible for the observed relaxor type ferroelectricity in this system. Across the temperature where anomaly in dielectric permittivity and change in structure was observed, appearance of peak in pyroelectric current data, further supports this finding. These experimental results provide evidences of the presence of relaxor type ferroelectricity in CCO system near room temperature, which can open up a route for search of room temperature relaxor type ferroelectricity in other delafossites systems also.

**ACKNOWLEDGEMENTS**

Authors would like to thank Tapas Ganguli for encouragement and support. Authors would also like to acknowledge the support of Anil Kumar IIT Indore for carrying out Raman measurements.

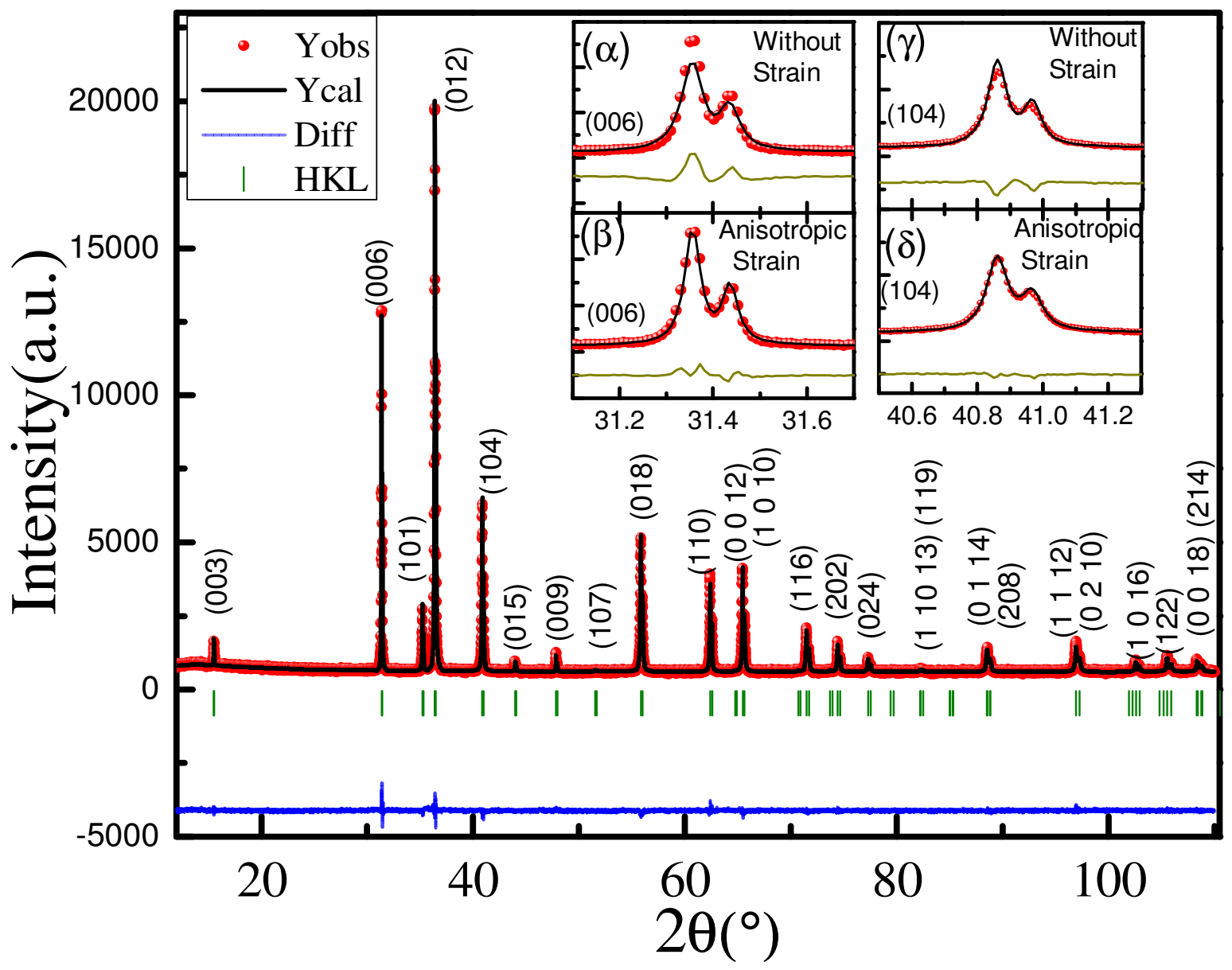

***Figure 1:****(a) Observed and calculated x-ray diffraction pattern for the synthesized CCO sample with space group R-3m*

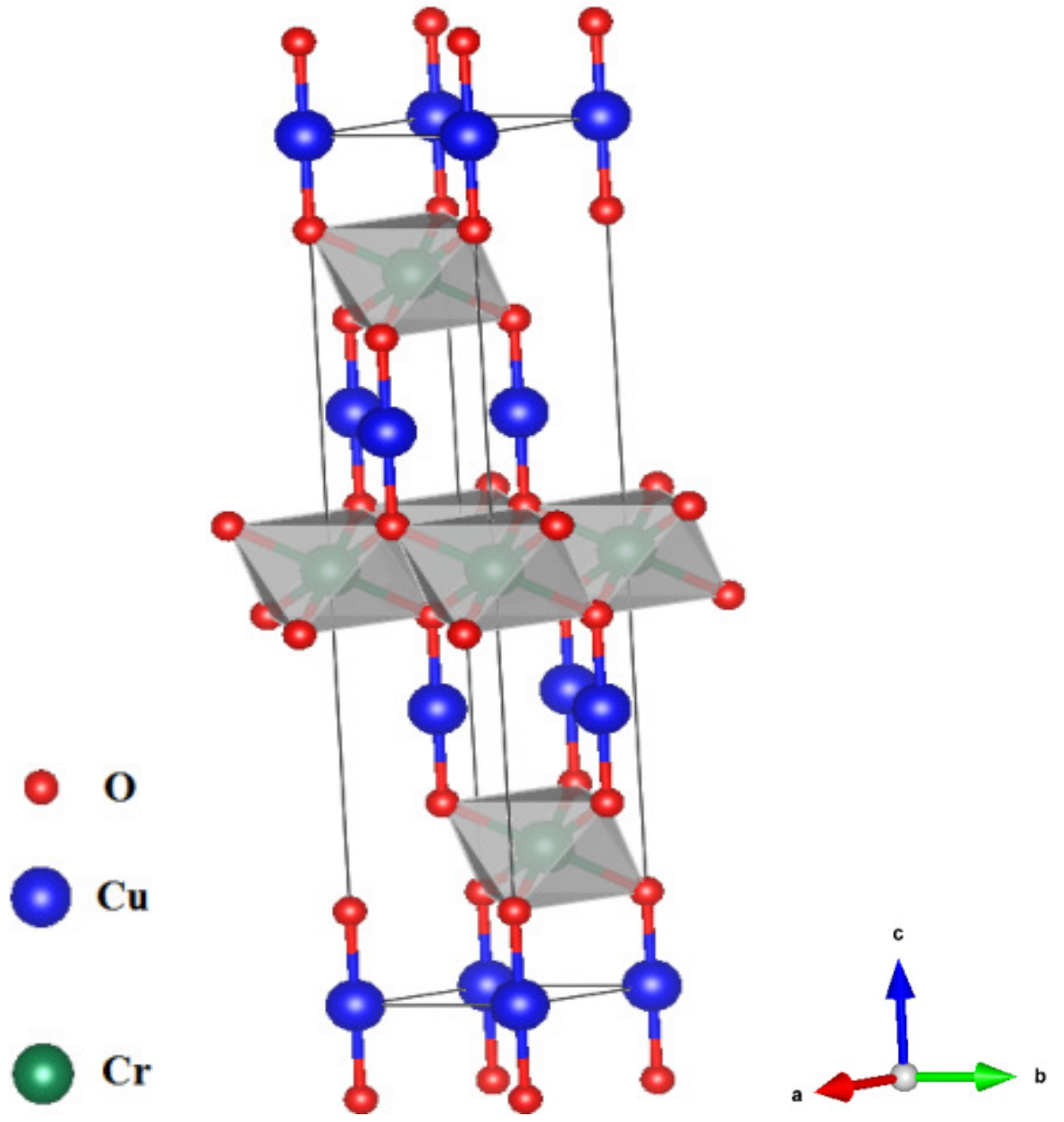

***Figure1** :(b) Schematic of crystal structure of CCO*

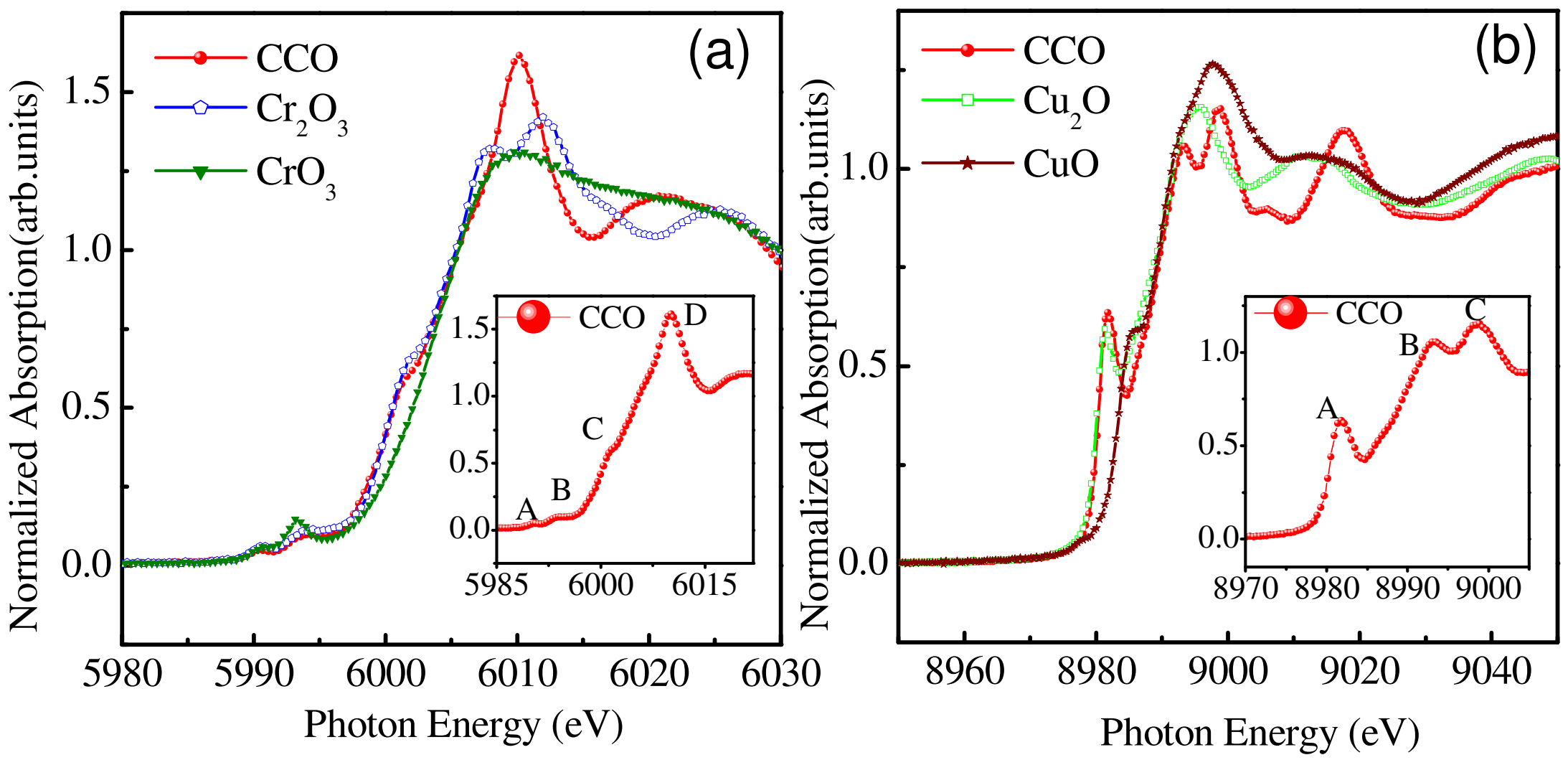

***Figure 2:****(a) Normalized XANES spectra at Cr K-edge of CCO sample along with* $Cr^{3+}$ *and* $Cr^{6+}$ *references. The inset shows the magnified Cr K-edge spectra of CCO (b) Normalized XANES spectra at Cu K-edge of CCO sample along with* $Cu^{1+}$ *and* $Cu^{2+}$ *references. The inset shows magnified Cu K- edge spectra of CCO.*

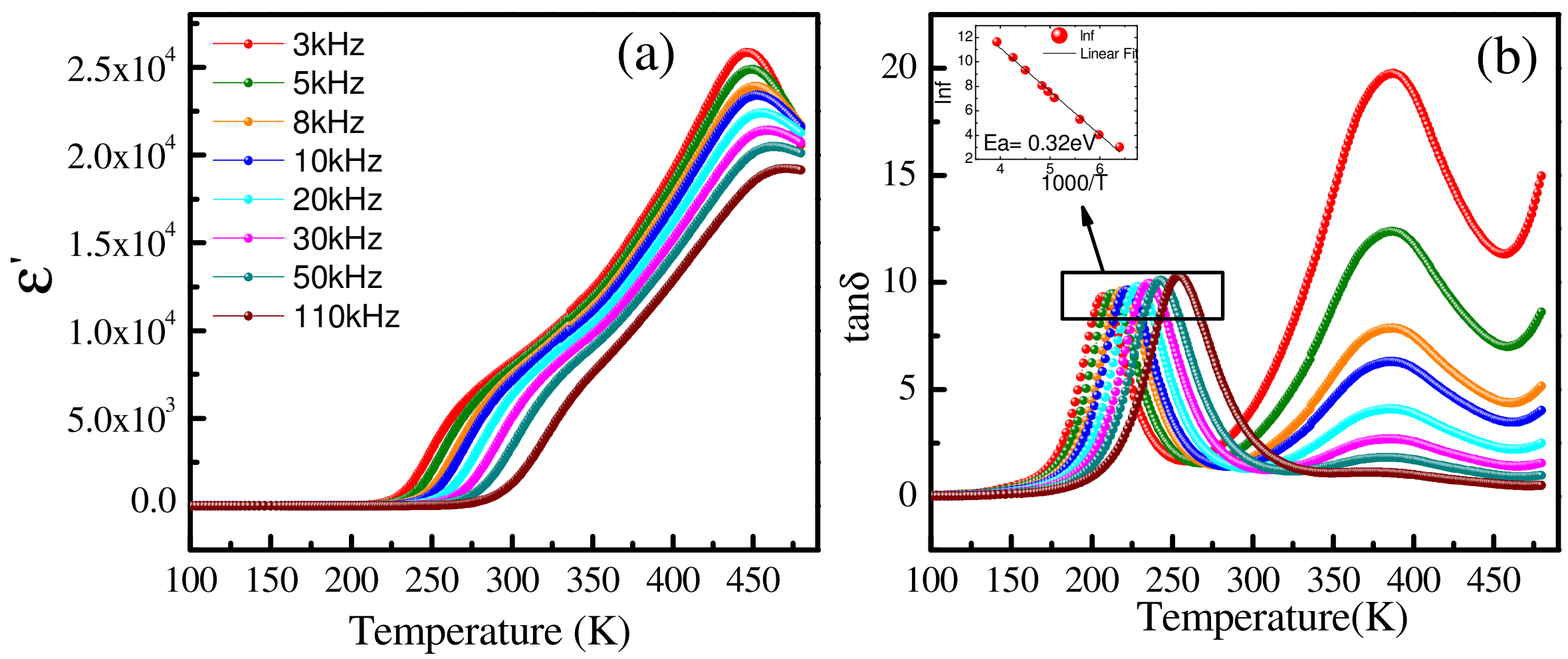

***Figure 3:****(a) Temperature dependence of dielectric permittivity at selected frequencies. Inset (α) shows variation of 1/ε' vs temperature at 3kHz (b) Temperature dependence of loss tangent at selected frequencies. Inset (β) shows Arrhenius plot corresponding to low temperature dielectric relaxation. Inset (γ) variation of loss tangent with temperature during heating and cooling at 5kHz.*

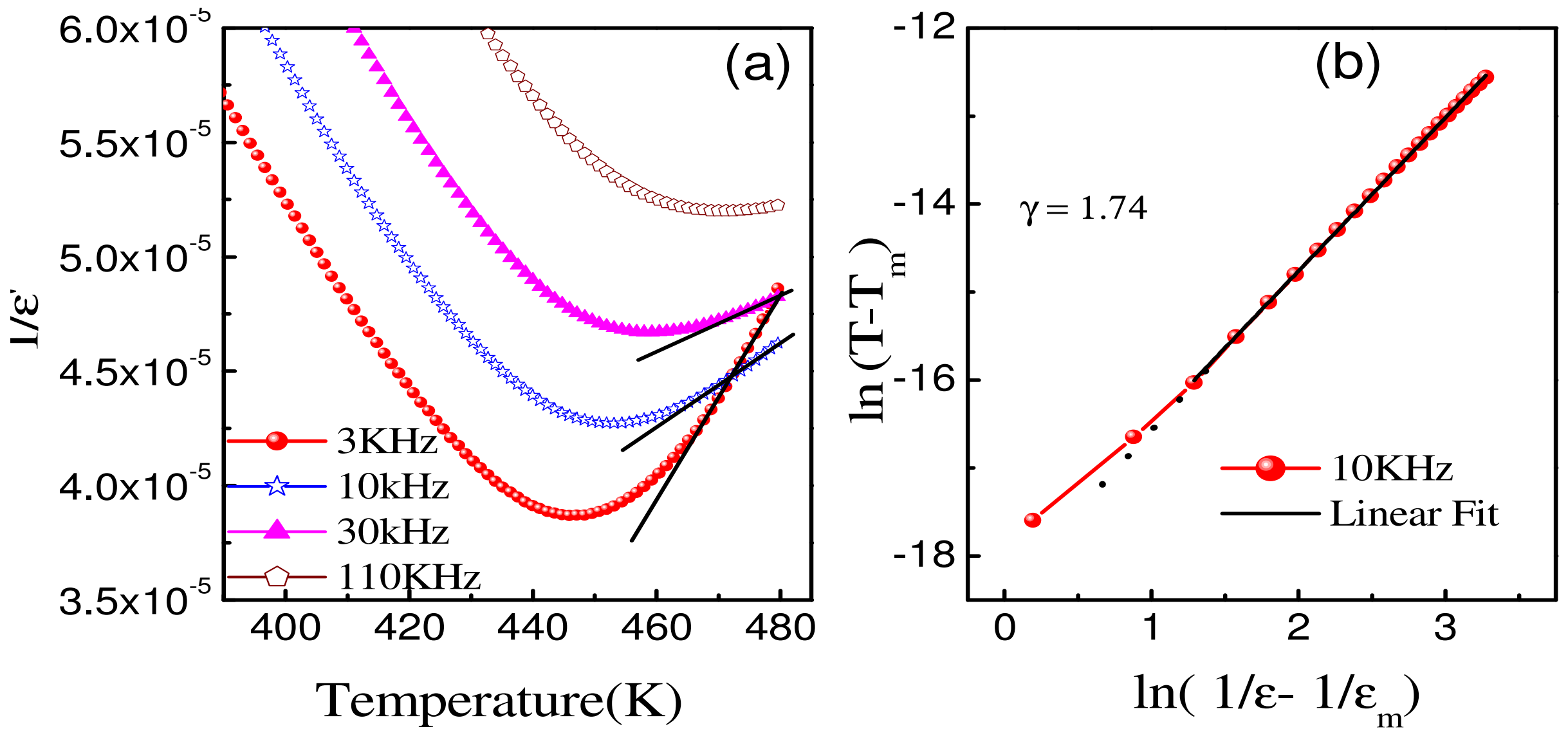

***Figure 4:*** *(a) Variation of inverse of permittivity with temperature (b) plot of ln (1/ε - 1/εm) vs ln (T-Tm)*

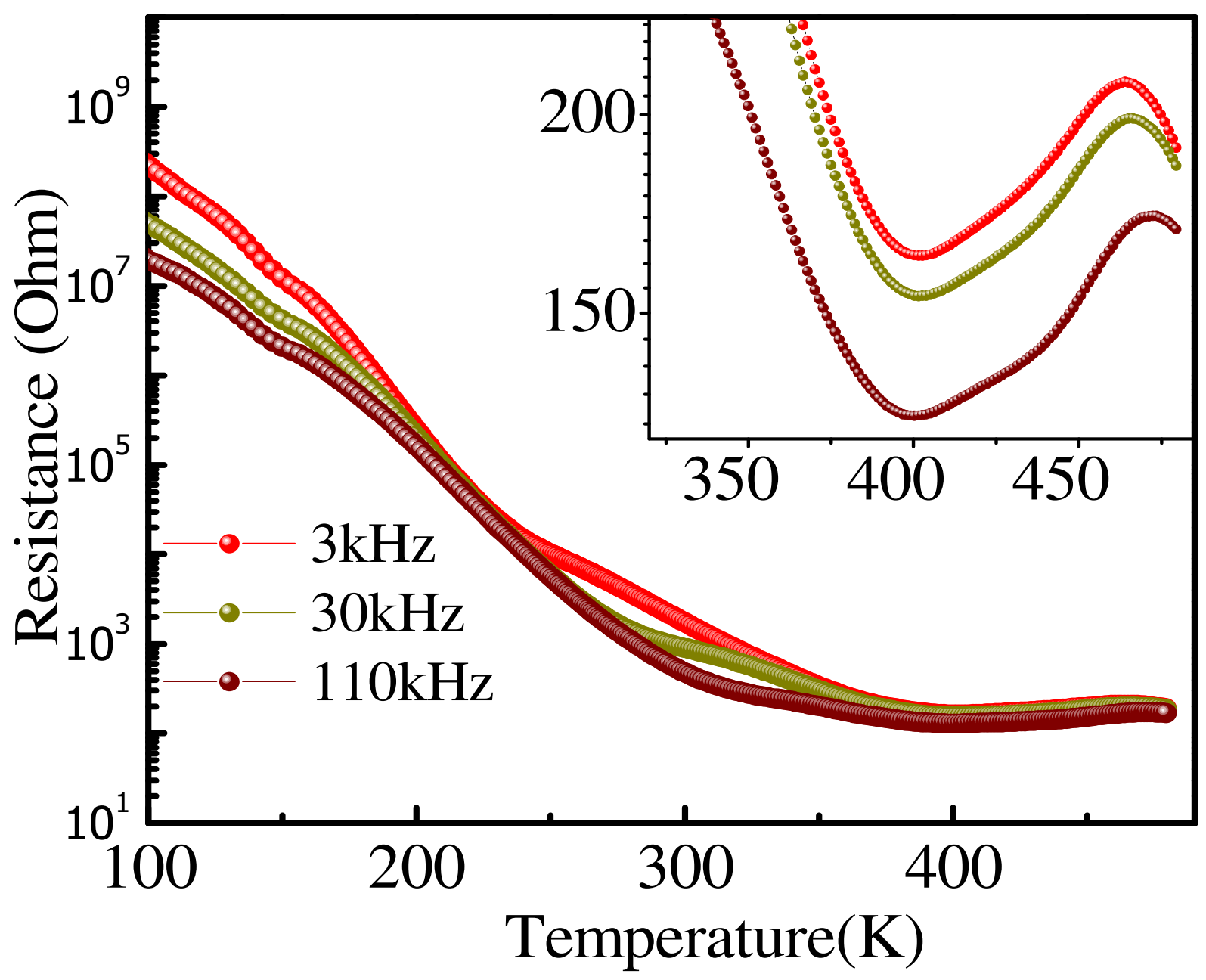

***Figure 5:*** *Temperature dependence of resistance at different frequencies.*

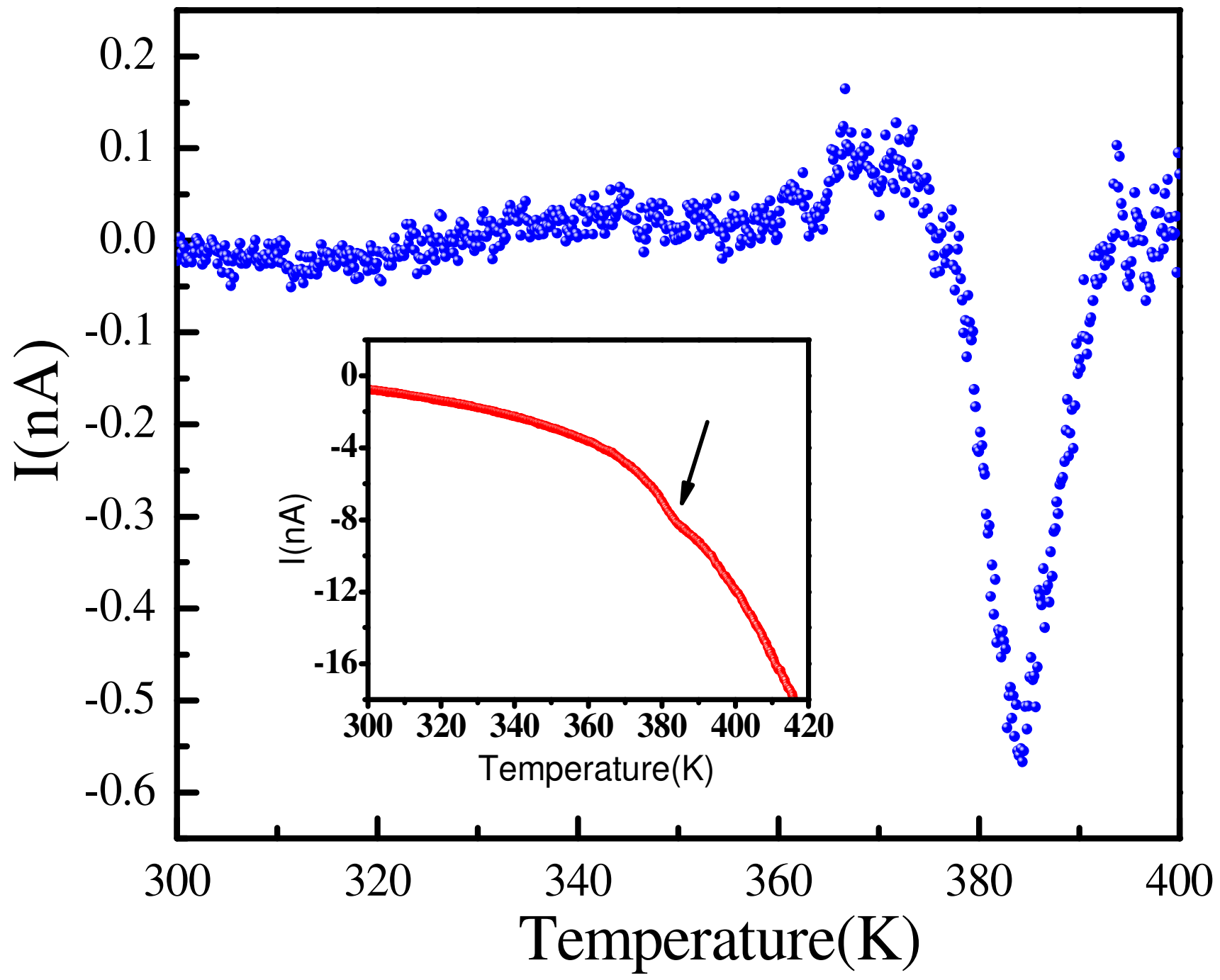

***Figure 6** : Pyroelectric current measured for poling field of 1.4kV/m after background subtraction as a function of temperature. Inset (a) Current measured as a function of temperature (b) polarization calculated from pyroelectric current*

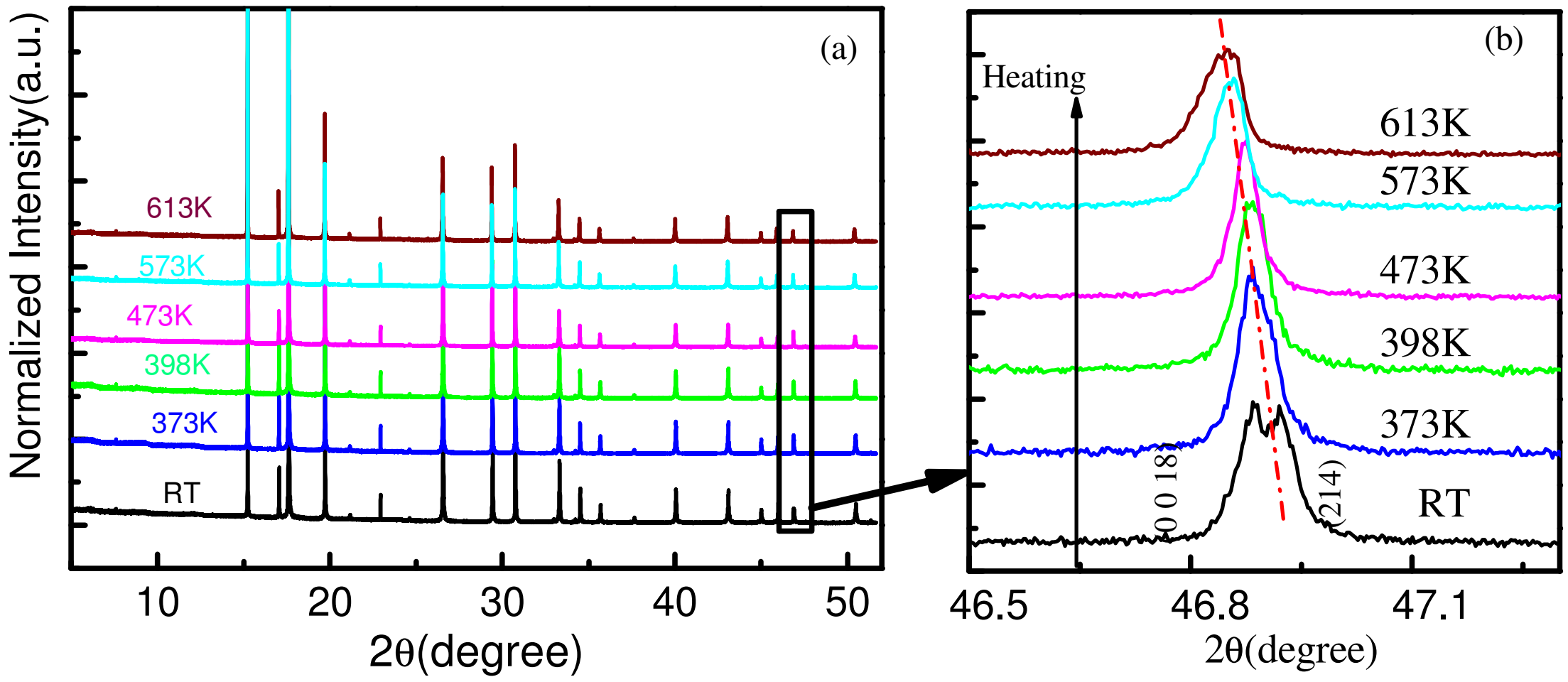

***Figure 7**: (a) Temperature dependent XRD pattern (b) merging of peaks during heating*

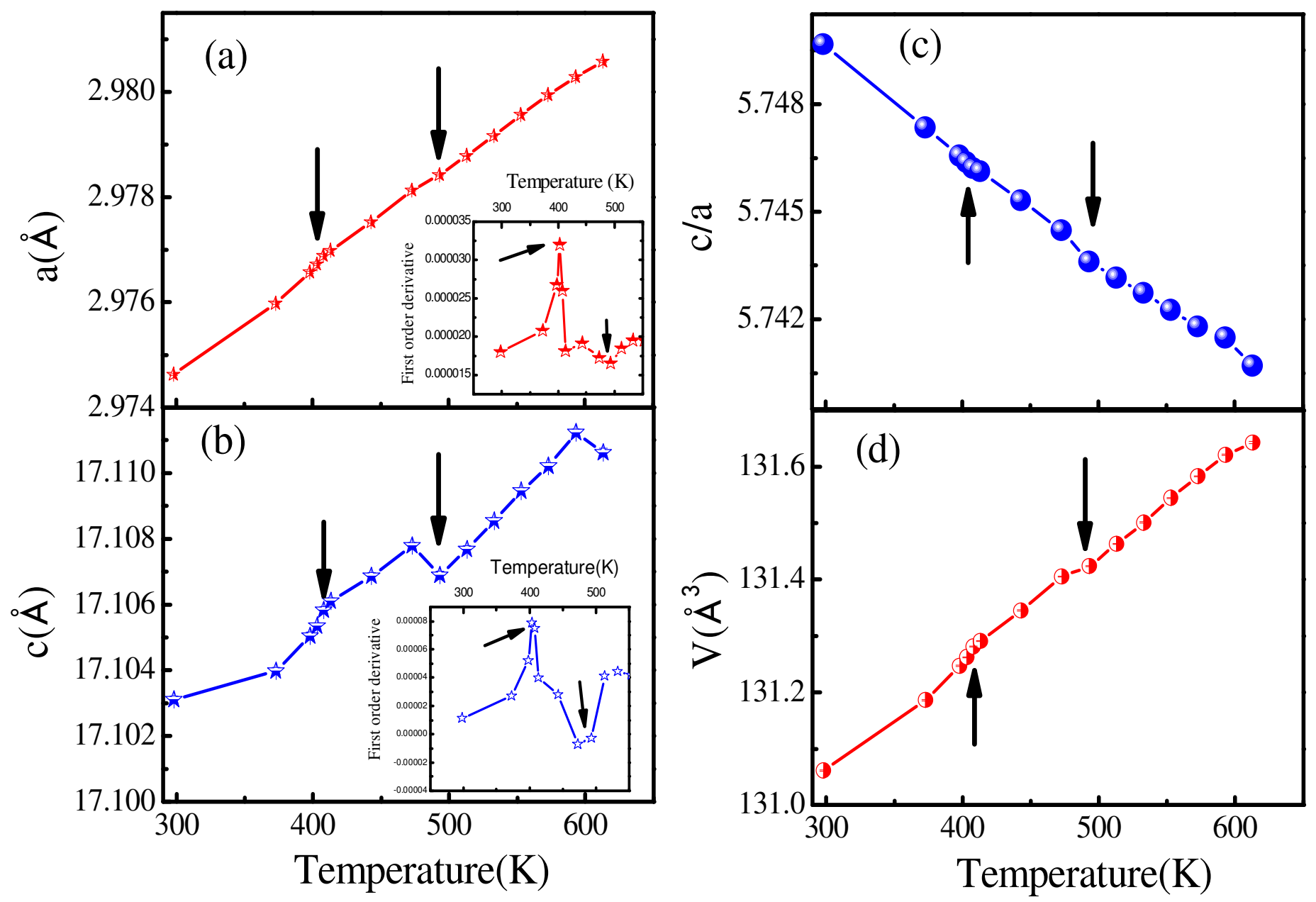

***Figure 8****: Temperature dependence of (a)(b) lattice parameter a and c (c) c/a ratio, (d) volume of unit cell, obtained from Rietveld refinement [Error bars are within size of legends*

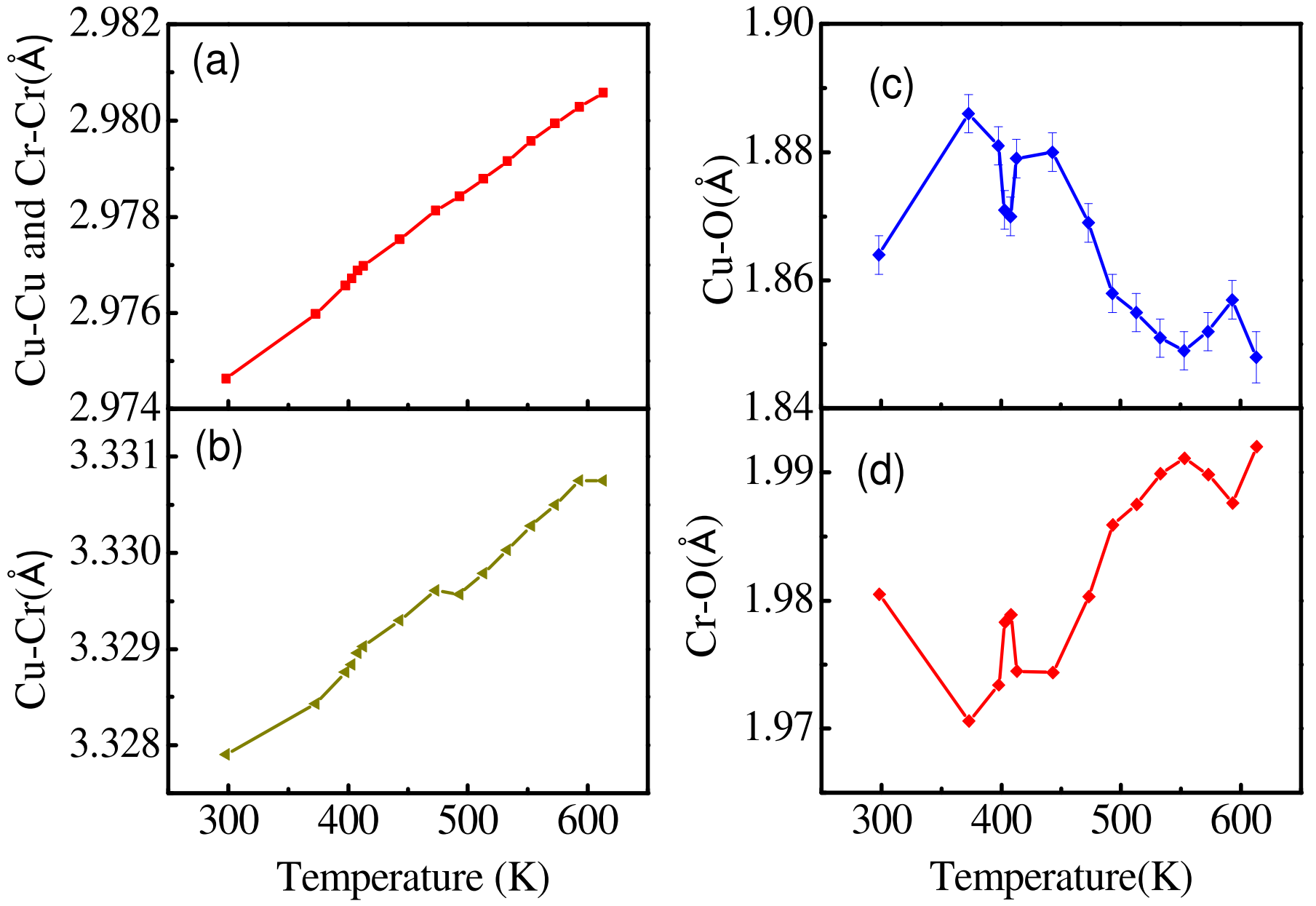

***Figure 9:*** *Temperature dependence of bond lengths (a) Cu-Cu and Cr-Cr (b) Cu-Cr (c) Cu-O (d)Cr-O, obtained from Rietveld refinement*

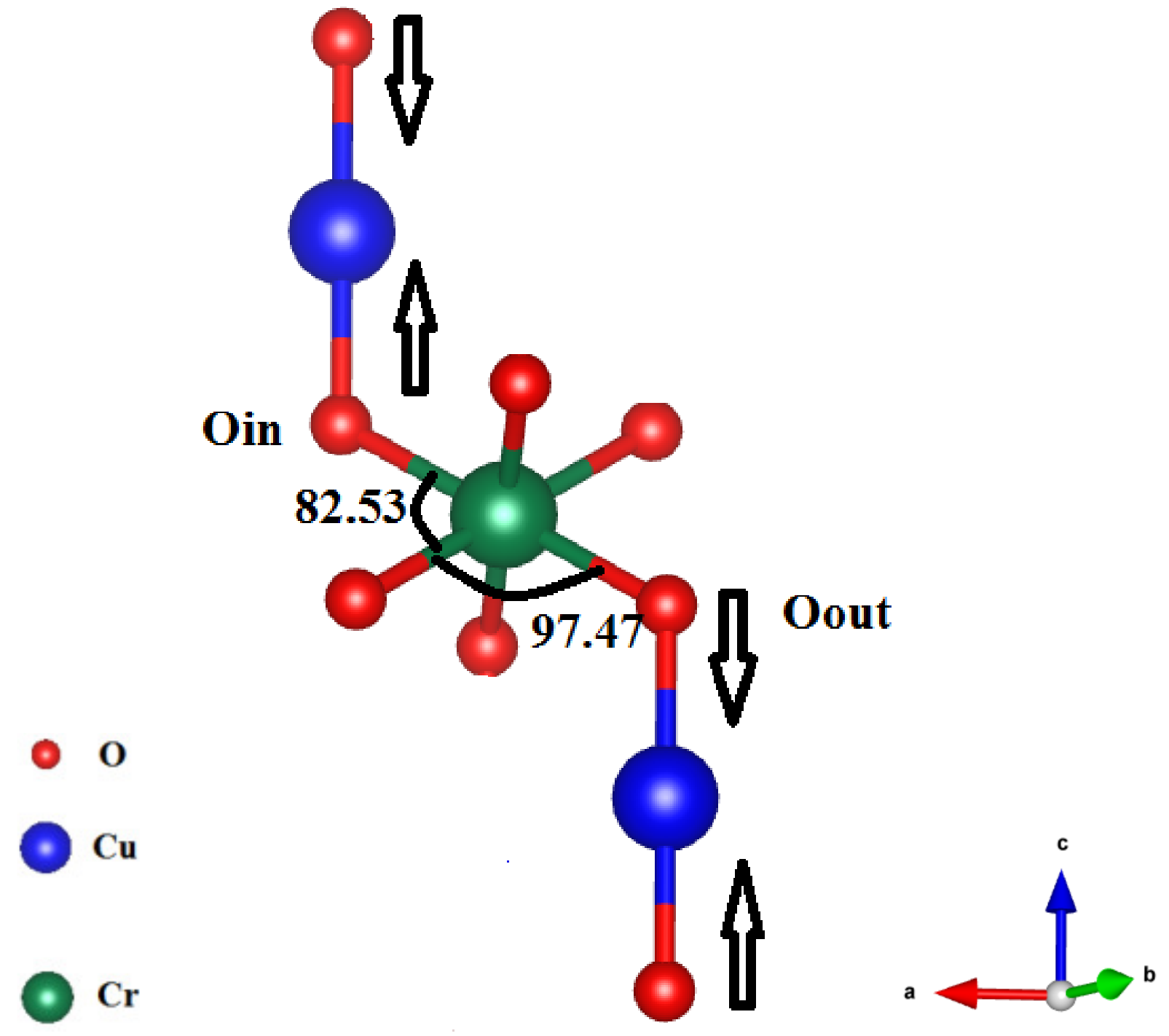

***Figure 10:*** *Schematic representation of octahedra $CrO_6$, ($O_{in}$ and $O_{out}$ represent two oxygen atoms at different distances)*

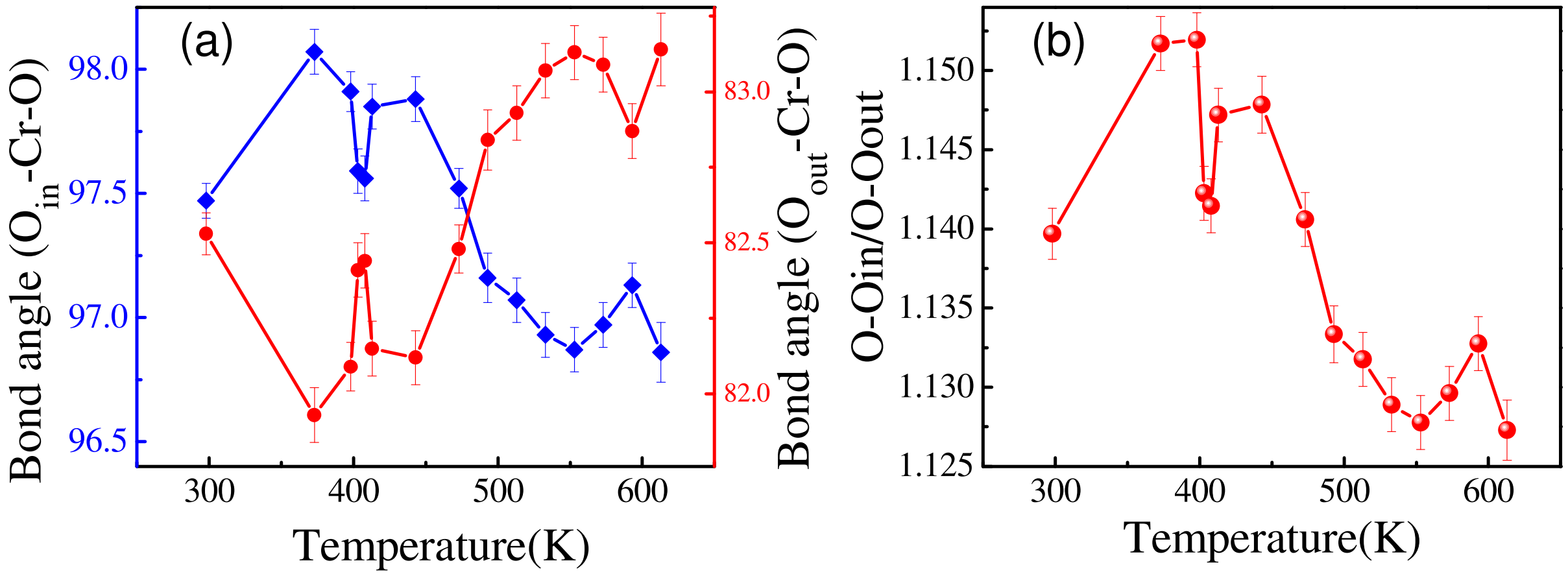

***Figure 11:*** *(a)Temperature dependence of bond angles $O_{in}$-Cr-O and $O_{out}$-Cr-O obtained from Rietveld refinement (b) Temperature dependence of degree of distortion of $CrO_6$ octahedra estimated from the ratio O-$O_{in}$/O-$O_{out}$*

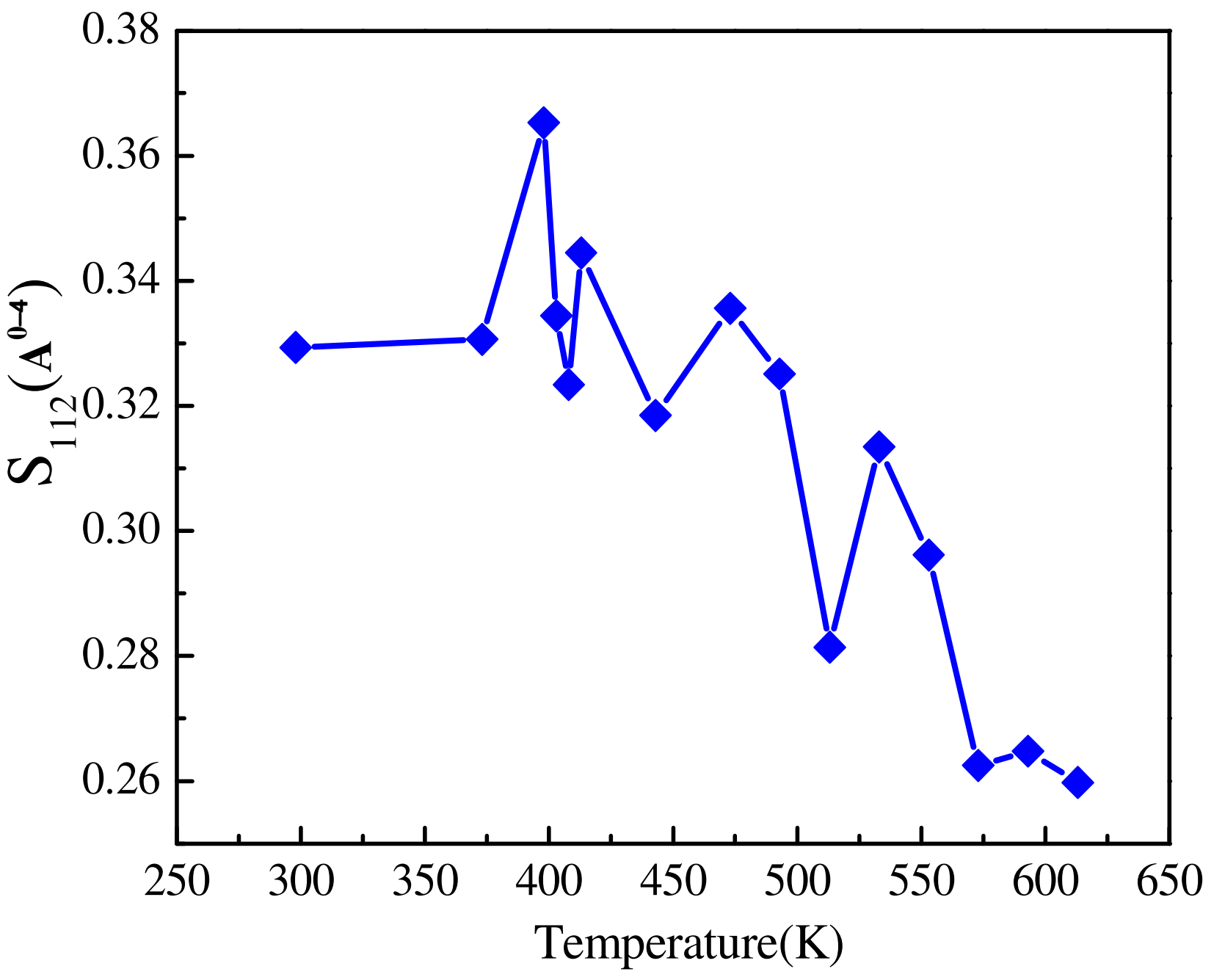

***Figure 12:*** *Temperature dependence of anisotropic broadening parameter $S_{112}$ ($Å^{-4}$) determined through Rietveld refinement*

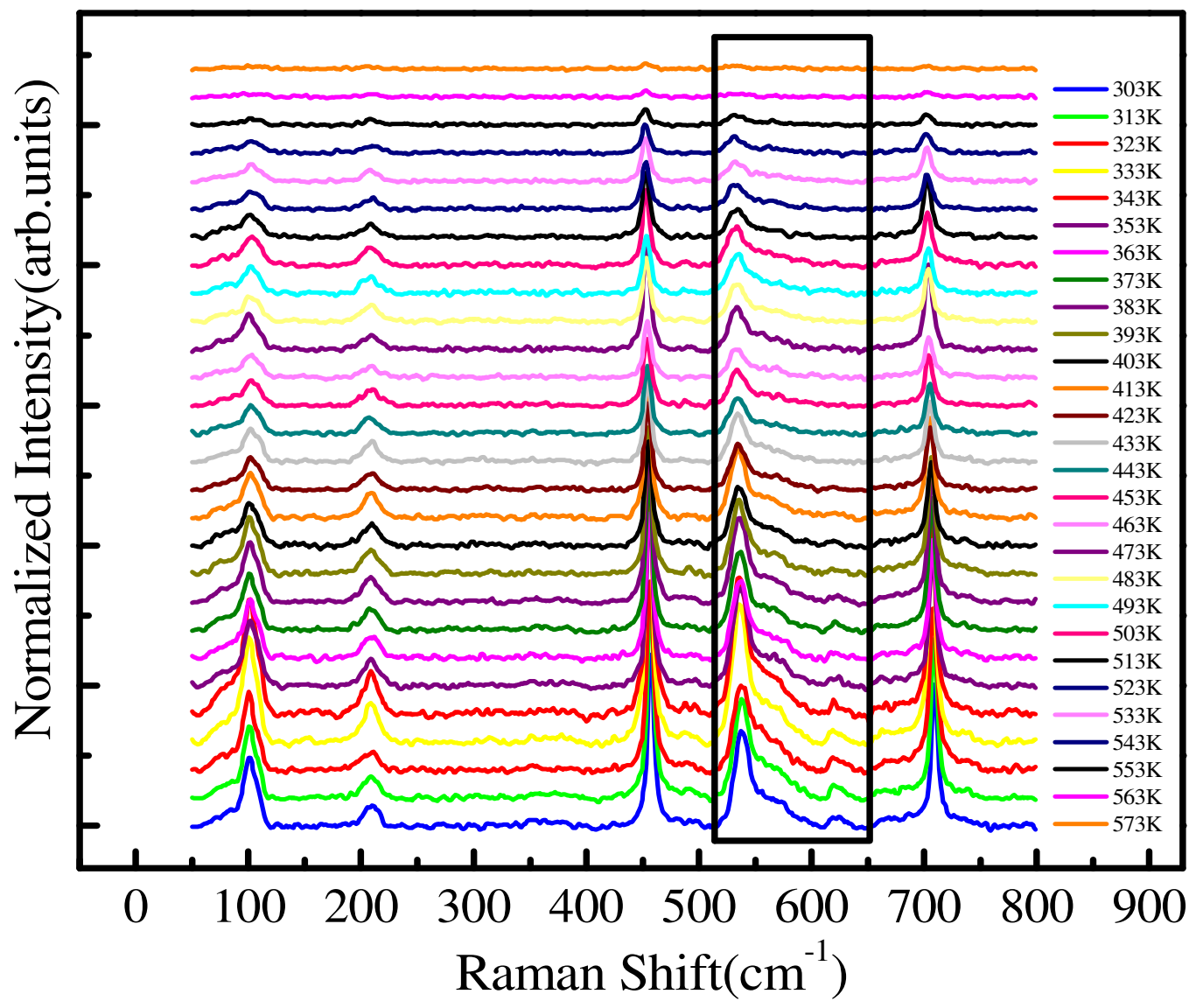

***Figure 13:*** *Raman spectrum of CCO in the temperature range of 303K to 573K*